\documentclass[a4paper,fleqn,usenatbib]{mnras}
\usepackage{newtxtext,newtxmath}
\usepackage{ae,aecompl}
\usepackage{graphicx}	
\usepackage{amsmath}	
\usepackage{amssymb}	
\usepackage[applemac]{inputenc}
\usepackage[T1]{fontenc}
\usepackage{lmodern}
\usepackage{textcomp}
\usepackage{hyperref}
\usepackage{natbib}
\usepackage[english]{babel}
\usepackage{fixltx2e}
\usepackage{subfigure}
\usepackage[labelfont=bf]{caption}
\usepackage{times}

\usepackage{fancyhdr}

\usepackage{color}
\newcommand{\silvia}[1]{{\leavevmode\color{black}#1}}

\def\be{\begin{equation}}
\def\ee{\end{equation}}
\def\ba{\begin{array}{lll}}
\def\ea{\end{array}}
\def\beq{\begin{eqnarray}}
\def\eeq{\end{eqnarray}}

\vspace{-2cm}
\title[Search for high-energy neutrinos from bright GRBs]{Search for high-energy neutrinos from bright GRBs with ANTARES}

\vspace{-2cm}
\author[The ANTARES Collaboration]{
\normalsize{A.~Albert$^{a}$, 
M.~Andr\'e$^{b}$, 
M.~Anghinolfi$^{c}$, 
G.~Anton$^{d}$, 
M.~Ardid$^{e}$, 
J.-J.~Aubert$^{f}$, 
T.~Avgitas$^{g}$, 
B.~Baret$^{g}$,
J.~Barrios-Mart\'{\i}$^{h}$, 
S.~Basa$^{i}$
% ...
}
\vspace{0.15cm}
\\
$^{a}$\scriptsize{GRPHE - Universit\'e de Haute Alsace - Institut universitaire de technologie de Colmar, 34 rue du Grillenbreit BP 50568 - 68008 Colmar, France} \\
$^{b}$\scriptsize{Technical University of Catalonia, Laboratory of Applied Bioacoustics, Rambla Exposici\'o,08800 Vilanova i la Geltr\'u,Barcelona, Spain} \\
$^{c}$\scriptsize{INFN - Sezione di Genova, Via Dodecaneso 33, 16146 Genova, Italy}\\
$^{d}$\scriptsize{Friedrich-Alexander-Universit\"at Erlangen-N\"urnberg, Erlangen Centre for Astroparticle Physics, Erwin-Rommel-Str. 1, 91058 Erlangen, Germany} \\
$^{e}$\scriptsize{Institut d'Investigaci\'o per a la Gesti\'o Integrada de les Zones Costaneres (IGIC) - Universitat Polit\`ecnica de Val\`encia. C/  Paranimf 1 , 46730 Gandia, Spain}\\
$^{f}$\scriptsize{Aix-Marseille Universit\'e, CNRS/IN2P3, CPPM UMR 7346, 13288 Marseille, France}\\
$^{g}$\scriptsize{APC, Universit\'e Paris Diderot, CNRS/IN2P3, CEA/IRFU, Observatoire de Paris, Sorbonne Paris Cit\'e, 75205 Paris, France}\\
$^{h}$\scriptsize{IFIC - Instituto de F\'isica Corpuscular (CSIC - Universitat de Val\`encia) c/ Catedr\'atico Jos\'e Beltr\'an, 2 E-46980 Paterna, Valencia, Spain}\\
$^{i}$\scriptsize{LAM - Laboratoire d'Astrophysique de Marseille, P\^ole de l'\'Etoile Site de Ch\^ateau-Gombert, rue Fr\'ed\'eric Joliot-Curie 38,  13388 Marseille Cedex 13, France}\\
% ...
}

\date{Accepted 2017 April 07. Received 2017 March 31; in original form 2016 December 24}

\pubyear{2017}

\begin{document}
\label{firstpage}
\pagerange{\pageref{firstpage}--\pageref{lastpage}}
\maketitle

\begin{abstract}
Gamma-ray bursts are thought to be sites of hadronic acceleration, thus neutrinos are expected from the decay of charged particles, produced in p$\gamma$ interactions. The methods and results of a search for muon neutrinos in the data of the ANTARES neutrino telescope from four bright GRBs (GRB 080916C, GRB 110918A, GRB 130427A and GRB 130505A) observed between 2008 and 2013 are presented. Two scenarios of the fireball model have been investigated: the internal shock scenario, leading to the production of neutrinos with energies mainly above 100~TeV, and the photospheric scenario, characterised by a low-energy component in neutrino spectra due to the assumption of neutrino production closer to the central engine. Since no neutrino events have been detected in temporal and spatial coincidence with these bursts, upper limits at 90\% C.L. on the expected neutrino fluxes are derived. The non-detection allows for directly constraining the bulk Lorentz factor of the jet $\Gamma$ and the baryon loading $f_p$.
\end{abstract}



\section{Introduction}
\label{sec:Intro}
The existence of hadronic acceleration mechanisms in Gamma-Ray Bursts (GRBs) would be unambiguously proven by the identification of high-energy neutrinos in temporal and spatial coincidence with the prompt emission of the burst. The detection of a single neutrino event would allow to identify this type of sources as a candidate for the Ultra-High-Energy Cosmic Ray (UHECR) production, whose origin is still under investigation (\citealt{blasi}). In order to test different scenarios, including those in which GRBs are able to reproduce the magnitude of the UHECR flux observed on Earth \silvia{(see for instance \citealt{waxman95}, \citealt{vietri95}, \citealt{wang08}, \citealt{murase08} and \citealt{globus})}, a multi-messenger approach can be adopted. For this purpose, the search for a possible neutrino counterpart can be crucial. Indeed, neutrinos are ideal candidates in the search for distant astrophysical sources, as they are electrically neutral, stable and weakly interacting particles.  \\
GRBs are transient sources, which release energies between $10^{51}$ and $10^{54}$~ergs in a few seconds (see \citealt{piran}, \citealt{meszaros} and \citealt{zhangkumarnew} for detailed reviews). Such extremely energetic events are probably related to the formation of a black hole, through the collapse of a massive star or the merging of a binary system (\citealt{piran}). The origin of GRB prompt emission is still under debate: the current theoretical understanding concerning the production of the $\gamma$-ray spectrum observed in the majority of GRBs is referred to as the standard fireball model (\citealt{piran98}), which naturally produces a non-thermal spectrum. The generally accepted picture is the Internal Shock (IS) scenario (\citealt{IS1}, \citealt{IS2} and \citealt{IS3}); nevertheless, the Photospheric (PH) scenario has also been widely discussed in literature (\citealt{PH1}, \citealt{PH2}, \citealt{PH3} and \citealt{ZANG}). They both assume that internal shocks take place when a faster shell of plasma catches up with a slower one: such a mechanism dissipates a large fraction of the kinetic energy of the flow, provided that the internal engine is highly variable. A fraction of this energy is expected to be transferred to accelerated particles: acceleration takes place on a very short timescale at the shock front, leading particles to ultra-relativistic speeds. 
Accelerated electrons subsequently radiate a fraction of their energy through synchrotron and inverse Compton processes. This radiation field constitutes the target for photohadronic interactions: 
from the collision of accelerated protons with the dense radiation field of the jet, mesons are produced, which then decay, producing neutrinos and $\gamma$-rays. \\
The main channel goes through the production of the $\Delta^+$ and its subsequent decay into pions, according to:

\begin{equation}
\label{eq:deltaplus}
p+\gamma 
\xrightarrow[]{\Delta^+}
\left\{
\begin{array}{l}
p+\pi^0 \\
n+\pi^+ \\ 
\end{array}
\right.
\longrightarrow \left\{
\begin{array}{l}
\pi^0 \longrightarrow \gamma + \gamma \\
n \longrightarrow p+e^- + \overline{\nu}_e \\
\pi^+ \longrightarrow \mu^+ + \nu_{\mu} \\
\mu^+ \longrightarrow e^+ + \nu_e + \overline{\nu}_{\mu} \\
\end{array}
\right.
\end{equation}

\noindent
In this dense environment, also kaon contribution becomes relevant to $\gamma$-ray production, because of the energy losses before their decay, and to neutrino production, especially at high energies. The treatment of neutrino production models from the prompt emission of GRBs was first given by \citealt{waxman} and in more detail by \citealt{guetta}.  \\
ANTARES (\citealt{antares}) is the largest undersea neutrino telescope on the Northern hemisphere, sensitive to neutrinos mainly with energies above hundreds of GeV. It is an array of photo-multiplier tubes (PMTs), anchored at a depth of 2475~m in the Mediterranean Sea, offshore Toulon (France). Neutrinos are detected through the Cherenkov radiation induced by ultra-relativistic particles created from a neutrino interaction. Track-like signatures are provided by muons, mainly produced by charged-current $\nu_{\mu}$ interactions. Previous searches for neutrinos from GRBs with both the ANTARES (\citealt{antares0}, \citealt{antares1} and \citealt{Julia}) and IceCube (\citealt{icecubeOld}, \citealt{icecube}) detectors did not measure any significant excess of neutrino events over the expected background and have placed limits on GRB parameters. \silvia{However,} recent works \silvia{on the IS model} (\citealt{multiRegion} and \citealt{winter}) suggest a GRB multi-zone production model for both neutrinos, \silvia{gamma-rays} and cosmic rays, which significantly lowers the neutrino expected flux with respect to previous predictions, indicating that such a flux may have been overestimated in earlier works.
\\
In this paper, a search for astrophysical neutrinos from bright GRBs with ANTARES data is presented. Bright sources represent promising targets, assuming that the neutrino flux scales with the $\gamma$-ray flux. In Sec.~\ref{sec:selection}, four bright GRBs used in the search for neutrinos are introduced. Then, in Sec.~\ref{sec:isph}, the internal shock and photospheric scenarios of the fireball model are briefly reviewed and the corresponding neutrino flux expectations are presented. Since the predicted signals are expected in different energy ranges, the analyses are performed using different data samples and specific features, as reported in Sec.~\ref{sec:methods}, where the analysis methods are outlined. The results are discussed in Sec.~\ref{sec:results}. Because of the fact that no neutrino has been observed in coincidence with the GRBs, constraints on the parameter space of the models are given in Sec.~\ref{sec:GRBconstr}: such constraints are derived for each GRB individually. Finally, the implications of such results on models for GRB neutrino production are examined in Sec.~\ref{sec:concl}.

\section{GRB selection}
\label{sec:selection}
The search for point-like neutrino sources consists of the identification of an event excess over the expected background from a given position in the sky, where the source is located, as illustrated in \citealt{antaresps}. In the case of GRBs, since the detected $\gamma$-ray emission is limited in time, also a temporal coincidence is required. In this way, it is possible to reduce the background contribution. The flux of atmospheric muons from above the detector comprises the largest part of the background, with a flux several orders of magnitude larger than any expected signal. The shielding effect of the Earth is exploited applying a geometrical cut on the reconstructed direction of the muon tracks. Selecting only upward going particles, the contamination by the atmospheric muons is largely reduced: since muons cannot cross the entire Earth, this cut rejects all atmospheric muons except for a small contamination due to mis-reconstructed events. In the study of transient sources, the requirement of temporal and directional coincidence allows to relax the cuts such that the dominant component is composed by mis-reconstructed atmospheric muons and atmospheric neutrinos, which represent an irreducible background for the cosmic signal. Therefore, an extended likelihood method is used to distinguish among signal and background events. For the search and simulation of neutrino fluxes, the brightest GRBs observable with ANTARES between 2008 and 2013 and the required $\gamma$-ray parameters are selected as described in Sec.~\ref{sec:grbsel}. Both the theoretical IS and PH models have been used to predict neutrino fluxes.

\subsection{GRB and $\gamma$-ray parameter selection}
\label{sec:grbsel}
GRBs with high observed $\gamma$-ray fluence, namely bursts with $F_\gamma>1\times 10^{-4} \, \textrm{erg} \, \textrm{cm}^{-2}$ (the average value of fluence ranges between $10^{-6}$ and $10^{-5} \, \textrm{erg} \, \textrm{cm}^{-2}$), were selected. It is also required that the progenitors of such bursts have the redshift measured, in order to estimate their intrinsic luminosity, and that they were in the field of view of the ANTARES telescope at the trigger time, i.e. located below the horizon. 
Four bright GRBs fulfill these criteria: GRB 080916C, GRB 110918A, GRB 130427A and GRB 130505A. In order to compute neutrino spectra, some input parameters are needed. However, some of them, which mainly concern the mechanism through which the jet kinetic energy is converted into internal energy, cannot be directly inferred from measurements. As a consequence, default values are assigned to these inputs: the ratio $f_p$ between internal energy in protons and electrons (also called baryonic loading) is fixed to $f_p=10$; the fraction of internal energy in electrons $\epsilon_\textrm{e}$ and that in magnetic field $\epsilon_\textrm{B}$ are assumed equal  because of energy equipartition, with $\epsilon_\textrm{e}=\epsilon_\textrm{B}=0.1$; the average fraction of proton energy transferred to a pion is $\langle x_{p \rightarrow \pi} \rangle =0.2$; and the Lorentz factor of the overall jet, more commonly denoted as bulk Lorentz factor, is $\Gamma=316$. Also, when not explicitly mentioned, the minimum variability time is assumed to be $t_{\textrm{var}}=0.01$ s for long bursts: this parameter affects the evaluation of neutrino expectations, since it is directly related to the morphology of the internal source (\citealt{tvar}). Below the selection of the $\gamma$-ray parameters, as collected from the Gamma-ray Coordinate Network (GCN) Circular Archive\footnote{\url{http://gcn.gsfc.nasa.gov/gcn3_archive.html}} and reported in Tab.~\ref{tab:param}, is described and the search strategy applied burst per burst is presented.\\
\newline
 \noindent
\textbf{GRB 080916C} triggered $\gamma$-ray satellites at 00:12:46 UTC on September 16th, 2008, with a right ascension $\textrm{RA}=119.87^{\circ}$ and declination $\textrm{DEC}=-56.59^{\circ}$. In a joint \emph{Fermi} GBM and LAT analysis (\citealt{latgbm}) five time bins are defined, relying  on the $\gamma$-ray spectral parameters. The relevant parameters for each bin in the burst are reported in Tab.~\ref{tab:param}. In particular, in bin B a 3~GeV photon was detected, followed by a 13.2~GeV photon in bin D: such high-energy emissions could be an indication of the hadronic origin of the radiation (\citealt{asano}). Moreover, the redshift of the progenitor was identified at $z=4.35$, while a minimum variability time scale of $t_{\textrm{var}}=0.23$~s was obtained from its light curve. Since neutrino production is directly linked to the GRB activity periods, our time dependent search is optimised in each of the five time bins defined by  \emph{Fermi} GBM and LAT analysis. The model expectations are therefore computed in each time bin and these contributions are summed up in order to obtain the expected signal from the burst. \\

\noindent
\textbf{GRB 110918A} started at 21:26:57 UTC on September 18th, 2011, located at $\textrm{RA}=32.58^{\circ}$ and $\textrm{DEC}=-27.58^{\circ}$ with a redshift $z=0.98$. Its local position in the ANTARES sky at the trigger time implied that neutrinos traveled up to the detector crossing the Earth quite horizontally, so that a negligible effect can be attributed to the Earth-absorption; this fact, together with the burst proximity in redshift, makes GRB 110918A a very promising candidate for a neutrino search with our detector.  A time-dependent search is performed on this burst, based on data in three time bins given from the \emph{Konus-Wind} satellite (\citealt{frederiks}), as reported in Tab.~\ref{tab:param}. \citealt{frederiks} also estimate the minimum variability time $t_{\textrm{var}}=0.25$~s. \\

\noindent
\textbf{GRB 130427A} enlight up the $\gamma$-ray sky on April 27th, 2013, at 07:47:07 UTC. From this burst two high-energy photons, of 95~GeV and 73~GeV, were detected by the \emph{Fermi} LAT satellite (\citealt{lat130427a}). The source position was reconstructed at $\textrm{RA}=173.14^{\circ}$ and $\textrm{DEC}=27.71^{\circ}$ with a redshift $z=0.34$. Its minimum variability time was measured to be $t_{\textrm{var}}=0.04$~s. The \textit{Konus-Wind} Collaboration provided the time-dependent spectral parameters of the main emission episode. \silvia{For this burst, an absence of neutrinos associated to the electromagnetic emission was announced by IceCube\footnote{\url{http://gcn.gsfc.nasa.gov/gcn3/14520.gcn3}} and then discussed in \citealt{gao}.}\\ 

\noindent
\textbf{GRB 130505A} happened on May 13th, 2013, at 08:22:27 UTC and $\textrm{RA}=137.06^{\circ}$, $\textrm{DEC}=17.49^{\circ}$ at $z=2.27$. Since the light curve of this burst is characterised by a main emission episode, a time average search was performed, relying on the spectral parameters released by \emph{Konus-Wind} on the GCN. For this burst, the default value of $t_{\textrm{var}}$ will be used in the following.

\tabcolsep=0.11cm
\begin{table*}
\caption{$\gamma$-ray parameters of each burst as detected from satellites (or, when not measured, assumed as default and marked with a *). Name of the burst; position in equatorial coordinates RA and DEC; time bin in case of time-dependent analysis; duration T; fluence $F_{\gamma}$ (measured in the energy range from 20~keV to 2~MeV for GRB 080916C and from 20~keV to 10~MeV for the others); low-energy spectral index $\alpha$, high-energy spectral index $\beta$ and peak energy $E_\gamma$ of a Band spectrum (\protect\citealt{band}); redshift $z$; minimum variability time $t_{\textrm{var}}$.}
\label{tab:param}
\centering 
\vspace{0.1cm}
\begin{tabular}{ccccccccccc}
\hline
NAME & RA & DEC & BIN & T & $F_{\gamma}$ & $\alpha$ & $\beta$ & $E_{\gamma}$ & $z$ & $t_{\textrm{var}}$\\
 & $(^{\circ})$ & $(^{\circ})$ & & (s) & ($10^{-4}$ erg/cm$^2$) & & & (keV) & & (s) \\
\hline
GRB 080916C & $119.87$ & $-56.59$ & A & $3.6$ & $0.15$ & $-0.58$ & $-2.63$ & $440$ & $4.35$ & $0.23$\\
" & " & " &  B & $4.1$ & $0.21$ & $-1.02$ & $-2.21$ & $1170$ & " & " \\
" & " & " & C & $48.2$ & $0.16$ & $-1.02$ & $-2.16$ & $490$ & " & " \\
" & " & " & D & $38.9$ & $0.53$ & $-0.92$ & $-2.22$ & $400$ & " & " \\
" & " & " & E & $46.1$ & $0.11$ & $-1.05$ & $-2.16$ & $230$ & " & " \\
GRB 110918A & $32.58$ & $-27.58$ & A & $2.3$ & $4.03$ & $-1.95$ & $-2.41$ & $990$ & $0.98$ & $0.25$ \\
" & " & " & B & $11.0$ & $2.06$ & $-1.00$ & $-2.60$ & $250$ & " & " \\
" & " & " & C & $15.1$ & $1.57$ & $-1.20$ & $-3.30$ & $78$ & " & " \\
GRB 130427A & $173.14$ & $27.71$ & - & $18.7$ & $26.8$ & $-0.96$ & $-4.14$ & $1028$ & $0.34$ & $0.04$ \\
GRB 130505A & $137.06$ & $17.49$ &  - & $7.0$ & $3.13$ & $-0.69$ & $-2.03$ & $631$ & $2.27$ & $0.01$* \\
\hline
\end{tabular}
\end{table*}

\section{The Internal Shock and the Photospheric models}
\label{sec:isph}
In GRB models, neutrinos are produced from the interaction between the accelerated protons and the jet radiation field. The predicted observable neutrino flux follows the primary spectrum; since both the internal shock and the photospheric models assume a differential energy spectrum for protons in the form of an unbroken power law with spectral index  $s=-2$ (according to the \silvia{standard diffusive shock} acceleration mechanism in the test-particle regime), also the energy of neutrinos will be distributed according to a power law spectrum. Two breaks are expected to modify the simple power law behavior of neutrino energy distribution, both due to synchrotron cooling of charged particles. The former reflects the break in the photon spectrum due to energy losses of accelerated electrons: it directly affects the neutrino spectrum since neutrinos result from photo-production processes. The latter break is due to the synchrotron losses from secondary mesons. The main difference between the two scenarios is the radius at which acceleration takes place, since it affects the optical depth $\tau_{p \gamma}$ of p$\gamma$ interaction (\citealt{waxman} and \citealt{ZANG}):

\begin{equation}
\tau_{p \gamma} = 0.8 \left( \frac{\textrm{R}}{10^{14}\textrm{cm}} \right)^{-1}  \left( \frac{\Gamma}{10^{2.5}} \right)^{-2}  \left( \frac{\textrm{E}_{\gamma}}{1\textrm{MeV}} \right)^{-1} \left( \frac{\textrm{L}_{\textrm{iso}}}{10^{52}\textrm{erg/s}} \right) 
\label{eq:opticalDepth}
\end{equation}

\noindent
where $\Gamma$ is the bulk Lorentz factor, R is the distance between the central engine and the neutrino production site which defines the fireball radius, $\textrm{L}_{\textrm{iso}}$ is the isotropic $\gamma$-ray luminosity of the burst and $\textrm{E}_{\gamma}$ is the energy at which the $\gamma$-ray spectrum has a break (of the order of 100 keV typically). In the IS scenario the radius of collision is (\silvia{\citealt{multiRegion}}):

\begin{equation}
\textrm{R}_{\textrm{IS}} = \silvia{2} \frac{c  t_{\textrm{var}}}{1+z} \, \Gamma^2 \sim 10^{13}  \, \left( \frac{ t_{\textrm{var}}}{0.01~\textrm{s}} \right) \left( \frac{\Gamma}{10^{2.5}} \right)^2  \left( \frac{1+2.15}{1+z} \right)~\textrm{cm}
\label{eq:radiusIS}
\end{equation}

\noindent
where $c$ is the light speed, $t_{\textrm{var}}$ is the minimum variability time scale observed in the light curve of the burst and $z$ is its redshift. The PH scenario predicts that particle acceleration occurs at a radius $\textrm{R}_{\textrm{PH}}$ (\citealt{zhangkumarnew}) in such a way that $\gamma$-rays are unable to escape due to high optical depth of electron-photon scattering (defined through the Thomson cross section $\sigma_T$):

\begin{equation}
\textrm{R}_{\textrm{PH}} = \frac{\textrm{L}_{\textrm{iso}} \sigma_T}{8\pi m_p c^3} \Gamma^{-3} \sim 10^{11} \, \left( \frac{\textrm{L}_{\textrm{iso}}}{10^{52} \textrm{erg/s}} \right)  \left( \frac{\Gamma}{10^{2.5}} \right)^{-3} \, \textrm{cm}  
\label{eq:radiusPH}
\end{equation}
\silvia{The photospheric model considers a baryonic dominated outflow, given the presence of the proton mass $m_p$ in Eq.~\ref{eq:radiusPH}: this assumption justifies the choice of the standard value $f_p=10$ for the baryonic loading in the prediction of neutrino spectra. Other formulations of the PH model exist in literature: \citealt{murasePRD} supposed that jets might also be dominated by pairs. In this case, the energy is mainly carried by radiation and a generally lower baryonic loading is assumed ($f_p \simeq 1$). Outflows with a huge neutron loading could also be considered, as in \citealt{murasePRL}: neutrinos in the energy range of tens of GeV are expected in this model, which makes these searches challenging for high-energy neutrino telescopes.} \\
For characteristic values of GRB parameters, Eq.~\ref{eq:radiusIS} and Eq.~\ref{eq:radiusPH} give $\textrm{R}_{\textrm{PH}}<\textrm{R}_{\textrm{IS}}$: $\tau_{p \gamma}$ in the PH model is enhanced by a factor $\textrm{R}_{\textrm{IS}}/\textrm{R}_{\textrm{PH}}$ compared to the IS model (see Eq.~\ref{eq:opticalDepth}). Consequently, the neutrino production is more efficient in a dissipative photosphere than in standard internal shocks. Finally, as the neutrino energy breaks depend on the radius (\citealt{ZANG}), in such a way that increasing the collision radius moves neutrino energy breaks to higher energies, the resulting PH model produces neutrinos at lower energy $(100~\textrm{GeV}-10~\textrm{TeV})$ than in the IS model $(100~\textrm{TeV}-1~\textrm{EeV})$. Therefore, the neutrino signal predictions are very different between the two models, as shown in the following. \\

\subsection{Neutrino flux expectations}
\label{subsec:neuFlux}
In this section, the methods used for the computation of the expected neutrino fluxes from each GRB are presented: they rely on the event generator \textquoteleft Neutrinos from Cosmic Accelerators\textquoteright (NeuCosmA), described in \citealt{Hummer2010}, for the IS model case and on the analytical description from \citealt{ZANG} in the PH model case.

\subsubsection{Internal Shock Model Case}
Detailed calculations of the GRB neutrino spectra in the IS context are performed, through the numerical code NeuCosmA.
Based on SOPHIA (\citealt{Mucke2000}), it simulates the particle physics with a pre-defined proton and photon spectrum (here a GRB Band spectrum, \citealt{band}) and takes into account the full p$\gamma$ cross section \silvia{(firstly derived in \citealt{murase06})}, including not only the $\Delta^+$ resonance but also higher mass resonances and kaon production. This yields an additional high-energy component in the $\nu_\mu$ spectrum, typically at EeV energies. Moreover, it considers individual energy losses of secondary particles and neutrino oscillations during their propagation from the source to the Earth. The normalization of the neutrino spectrum is linearly scaled to the baryonic loading factor and to the per-burst $\gamma$-ray fluence. The algorithm produces the expected neutrino spectrum, assuming the measured values of the $\gamma$-ray parameters, as reported in Tab.~\ref{tab:param} for each emission episode of the bursts. The resulting muon neutrino spectra are given as solid lines in the left panel of Fig.~\ref{fig:Limit}. 

\subsubsection{Photospheric Model Case}
To compute the PH neutrino spectra the general formalism developed by \citealt{ZANG} was adopted, which adds a correction factor $f$ to the normalization to take into account the fact that only a fraction of the accelerated protons will produce neutrinos via p$\gamma$ interactions. \silvia{No $pp$ interaction is considered in the \cite{ZANG} model.} These fluxes are shown as solid lines in the right panel of Fig.~\ref{fig:Limit}. Because the energy range of interest for this search is below 10~TeV, special features that could offer a better ANTARES sensitivity in the lower energy range have been used in this analysis: a sample of unfiltered data, a low-energy optimised reconstruction algorithm and a directional filter, as described in Sec.~\ref{subsec:data}.

\section{Methodology}
\label{sec:methods}
Two different data samples are used in order to match the neutrino energy range expected from the two models, each with specific features concerning the track reconstruction algorithms, background evaluation and search time windows, as reported in Sec.~\ref{subsec:data}. The same optimisation method is used for both models and is described in Sec.~\ref{subsec:testStatistic}.

\subsection{Data samples and specific analysis features}
\label{subsec:data}
The ANTARES Data Acquisition (DAQ) system (\citealt{antaresDAQ}) is designed following the "all data to shore" concept: all photon signals are recorded above a threshold of 0.3 photo-electrons by the optical modules. They are then sent and buffered in the shore station where a filtering is performed. In some special cases, such as a GRB alert, the full unfiltered buffer can be saved on disk. The ANTARES detector receives the the GCN alert, which contains the position of the burst and its main features. In 90\% of the cases the delay between the detection of a GRB by the satellite and the time of the alert message distributed is below 200~s (with a typical delay around 10~s). The GRB unfiltered data sample also includes unfiltered data buffered before the alert message reception. The overall size of the unfiltered data sample is about 2 minutes, so that data cover the majority of the burst duration (\citealt{mieke}). 
To increase the sensitivity to low-energies, unfiltered data are used for the PH model, while filtered data are used for the IS one. The unfiltered data recorded are analysed with a dedicated algorithm, searching for space-time correlations restricted in a small search cone centered to the position of the considered GRB. A less strict filter condition with respect to the standard online triggers is applied. This algorithm yields more detected events in the target direction. A dedicated reconstruction algorithm (\citealt{GRID}), optimised for energies below 1~TeV, is also applied to this specific data set. Through these new features and following the same search method presented in Sec.~\ref{subsec:testStatistic}, but with a dedicated muon background estimation, the sensitivity improves by a factor of two at energies between 100~GeV and 1~TeV, where most of the neutrino flux is expected according to the PH model. The analysis performance is compatible with the one of the IS analysis at higher energies.

\subsection{Analysis method}
\label{subsec:testStatistic}
In order to simulate the per burst expected signal, the standard ANTARES Monte Carlo simulation chain has been used. It accurately describes the data taking conditions and the detector response during each GRB.
The background for each burst is evaluated with data: upgoing atmospheric neutrinos are the main background component, with a smaller contribution coming from mis-reconstructed downgoing atmospheric muons. The number of background events $\mu_b$ expected in a defined angular and temporal window around the burst location is therefore assumed to be known \emph{a priori}. The search time window in the IS analysis is chosen to be equal to each burst duration T (obtained as the sum of the time-bin durations) with a symmetric extension of 2 seconds. To be conservative, this extension is much larger than any effect due to the light propagation time from the satellite to our detector and to uncertainties in the DAQ system. In the PH case, instead, the time window depends on the unfiltered data buffer duration. Since GRBs are transient sources, the angular window of the search can be enlarged with respect to that normally used in a steady source search (\citealt{antaresps}): the search cone around the burst is fixed with an aperture equal to $10^{\circ}$. Given the short duration time window, this value still allows to have a rate of expected signal generally higher than the estimated background in the same search region, as will be shown later. \\
The analysis is optimised independently for each burst, as described in \citealt{Julia}, through the computation of pseudo-experiments with $n_{\textrm{tot}}$ total number of events, based on an extended maximum likelihood ratio test statistic $Q$ (\citealt{barlow}):

\begin{equation}
\label{eq:testStat}
Q = \max_{\mu^{\prime}_{s} \in [0;n_{\textrm{tot}}]}
\Bigl( \sum\limits_{i=1}^{n_{\textrm{tot}}} \log \frac{\mu^{\prime}_s S(\alpha_i)+\mu_b B(\alpha_i)}{\mu_b B(\alpha_i)}-\mu^{\prime}_s\Bigr)
\end{equation}

\noindent
where $\alpha_i$ is the angular distance between the GRB position and the reconstructed muon direction, $S(\alpha_i)$ is the signal probability density function, obtained from Monte Carlo simulations, and $B(\alpha_i)$ is the background probability density function, assumed flat in the solid angle of the cone. In order to extract the distribution of $Q$ as a function of the injected signals, more than $10^8$ pseudo-experiments have been performed. Signal and background events are randomly extracted from their normalised distributions and the test statistic evaluated, returning the estimated signal $\mu^{\prime}_s$ as the one maximising $Q$ . The significance of a measurement is given by its $p$-value\footnote{A Gaussian two-sided convention is applied, with a $3\sigma$ background rejection corresponding to a $p$-value of $p_{3\sigma}=2.7 \times 10^{-3}$.}, that is the probability of getting values for $Q$  at least as high as that observed if the background only hypothesis were true.\\
This procedure is repeated for different cut value of the track quality parameter (\citealt{antaresLambda}): the finally selected value for this parameter is the one that maximises the probability to observe an excess with a $p$-value lower than the pre-defined threshold at a given statistical accuracy, assuming the expected signal flux from the model.

\begin{figure*}
\begin{minipage}{.4\textwidth}
\hspace{-1.5cm}
\includegraphics[scale=0.48]{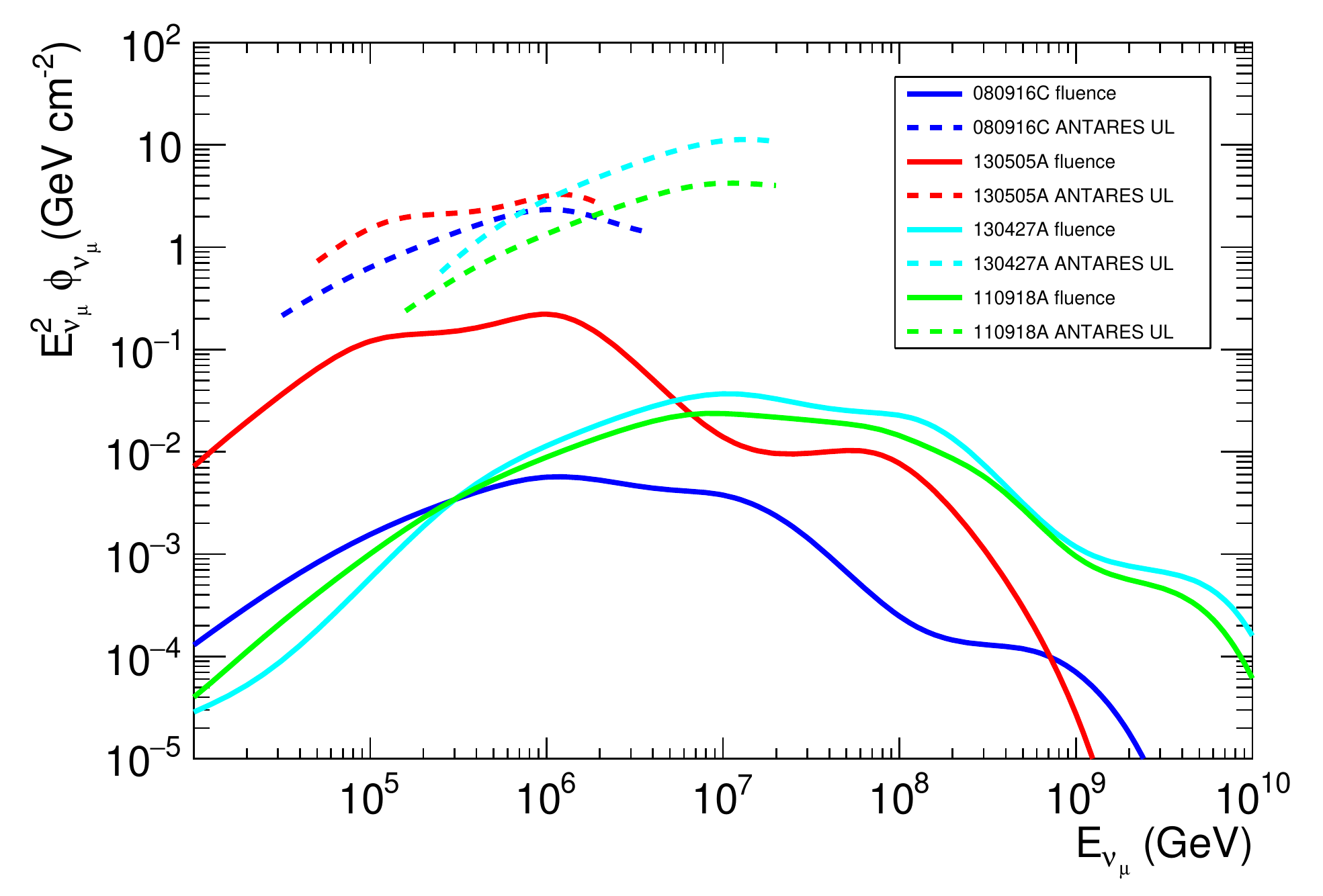}
\end{minipage}
\qquad \hspace{0.1cm}
\begin{minipage}{.4\textwidth}
\includegraphics[height=0.27\textheight, width=1.21\textwidth]{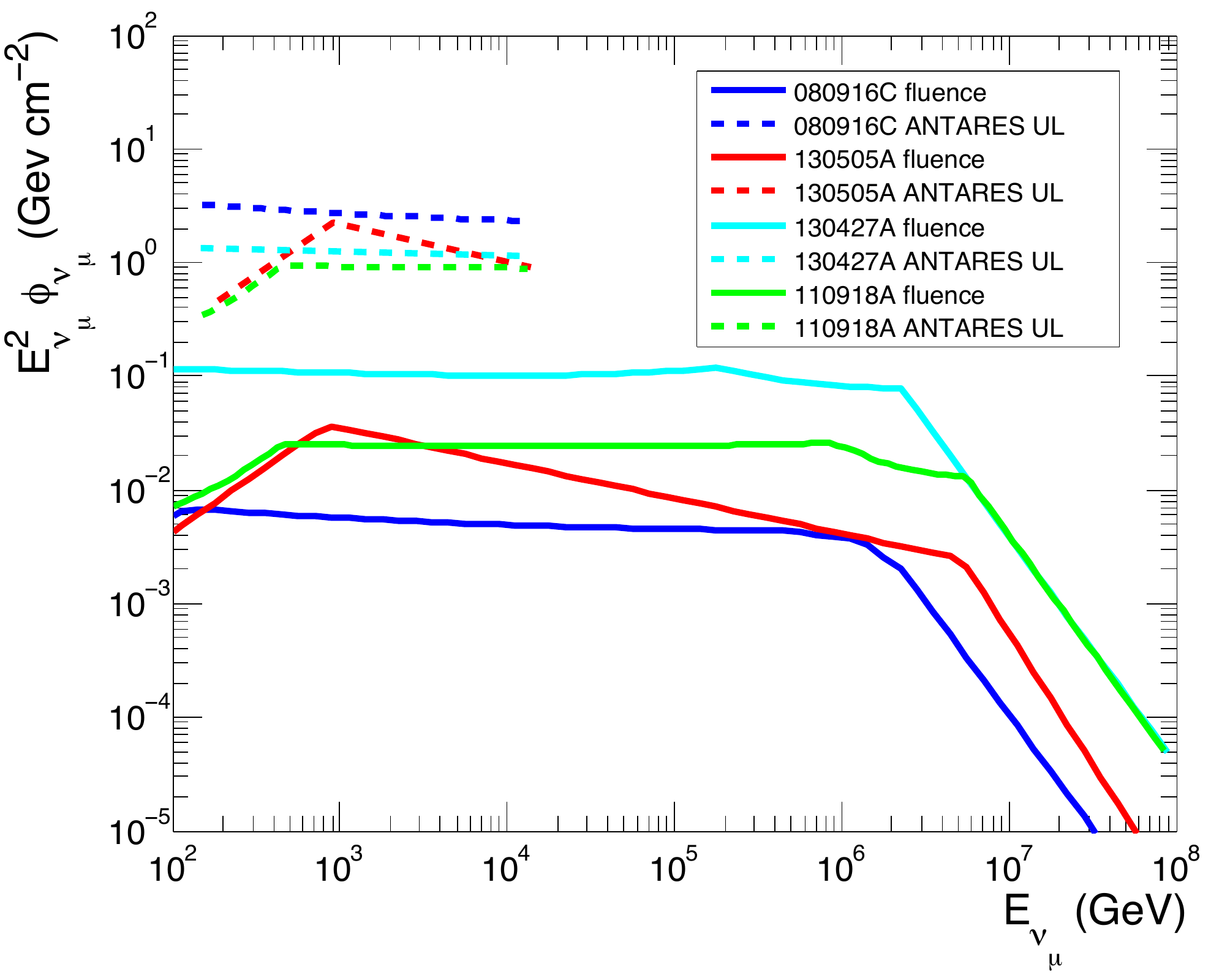}
\end{minipage}
\caption{Solid lines: expected $\nu_\mu+\overline{\nu}_{\mu}$ fluences. Dashed lines: ANTARES 90\% C.L. upper limits on the selected GRBs, in the energy band where 90\% of the signal is expected to be detected by ANTARES. {\it Left:} IS model prediction (NeuCosmA). {\it Right:}  PH model prediction.} 
\label{fig:Limit}
\end{figure*}

\section{Results}
\label{sec:results}
Both analyses are optimised for the track quality cut yielding the maximum detection probability for a 3$\sigma$ significance, with the background event rate $\mu_b$ evaluated as in \citealt{Julia}. The results of the optimised IS analyses on the four bursts are summarised in Tab.~\ref{tab:optimum}. From these results, it is evident that for three bursts (GRB 110918A, GRB 130427A and GRB 130505A) the estimated background $\mu_b$ is smaller than the expected signal $\mu_s$. \\
After the analyses have been optimised for each burst, the different track quality cuts have been applied. In the PH case, the strategy described in Sec.~\ref{subsec:data} was applied on the unfiltered data files recorded in coincidence with GRB 130427A and GRB 130505A (since for GRB 080916C and GRB 110918A unfiltered data were not available). No events have been detected in spatial and temporal coincidence with any of these bursts in any of the time windows selected for the searches. 90\% C.L. upper limits on the expected signal fluences are derived and reported in Fig.~\ref{fig:Limit}. Defining the differential neutrino fluence $\phi_{\nu}$, our limits are $E_{\nu}^2 \phi_{\nu}$ between about $[0.1-10]$~GeV/cm$^2$ for both models. Concerning the IS scenario, the closest upper limit to the expected flux is derived for GRB 130505A. This may also be related to the fact that it is the only burst of the sample for which the default value of minimum variability time scale has been used, because it was not directly measured. GRB 110918A and GRB 130427A give quite similar results: the better limit is on GRB 110918A, given the better effective area of the detector at the local position of this burst; the upper limit on GRB 080916C is on the other hand limited by its high redshift. For the PH scenario limits on the bursts for which unfiltered data were not available are obtained assuming no detection and using the optimised cuts of the IS analysis. \\
\silvia{The individual limits derived for these bright GRBs are consistent with the limits shown in previous ANTARES stacking searches (see Fig.~8b in \citealt{Julia}), which refer to the IS model only. In the standard approach $f_p$ and $\Gamma$ are assumed to be the same for all the stacked sources, respectively equal to $f_p=10$ and $\Gamma=316$: this assumption leads bright GRBs to be the main contributors to the total neutrino flux, even when numerous but fainter GRBs are added to the search.}

\begin{table}
\caption{Optimised 3$\sigma$ search for the four bursts, based on the IS model: columns report the optimised number of expected background and signal events, $\mu_b$ and $\mu_s$ respectively, and the probability to discover an excess (MDP) as predicted from the NeuCosmA model.}
\label{tab:optimum}
\centering 
\begin{tabular}[hb]{cccc}
\hline
NAME & $\mu_b$ & $\mu_s$ & MDP \\
\hline
GRB 080916C & $8.6 \times 10^{-3}$ & $1.8 \times 10^{-3}$ & $4.4  \times 10^{-3}$\\
GRB 110918A & $7.2 \times 10^{-3}$ & $1.3 \times 10^{-2}$ & $1.5 \times 10^{-2}$\\
GRB 130427A & $4.1 \times 10^{-3}$ & $7.5\times 10^{-3}$ & $ 8.7 \times 10^{-3}$ \\
GRB 130505A & $2.4 \times 10^{-3}$ & $1.6 \times 10^{-1}$ & $1.5 \times 10^{-1}$\\
\hline
\end{tabular}
\end{table}

\section{Constraints on GRB Models}
\label{sec:GRBconstr}
\begin{figure*}
\hspace{-1.7cm}
\begin{minipage}{.4\textwidth}
\includegraphics[width=1.25\textwidth]{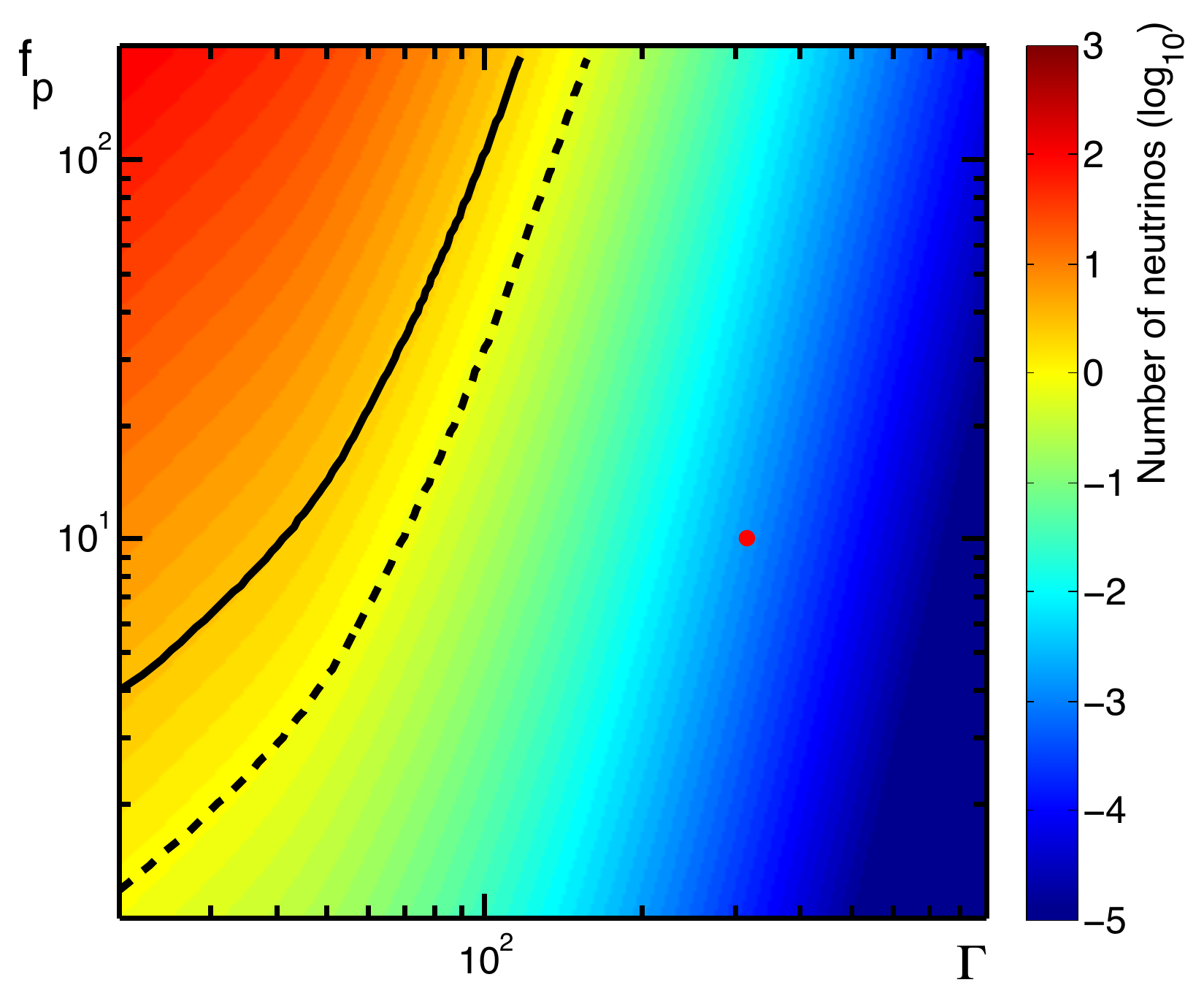}
\end{minipage}
\qquad \hspace{1.3cm}
\begin{minipage}{.4\textwidth}
\includegraphics[width=1.25\textwidth]{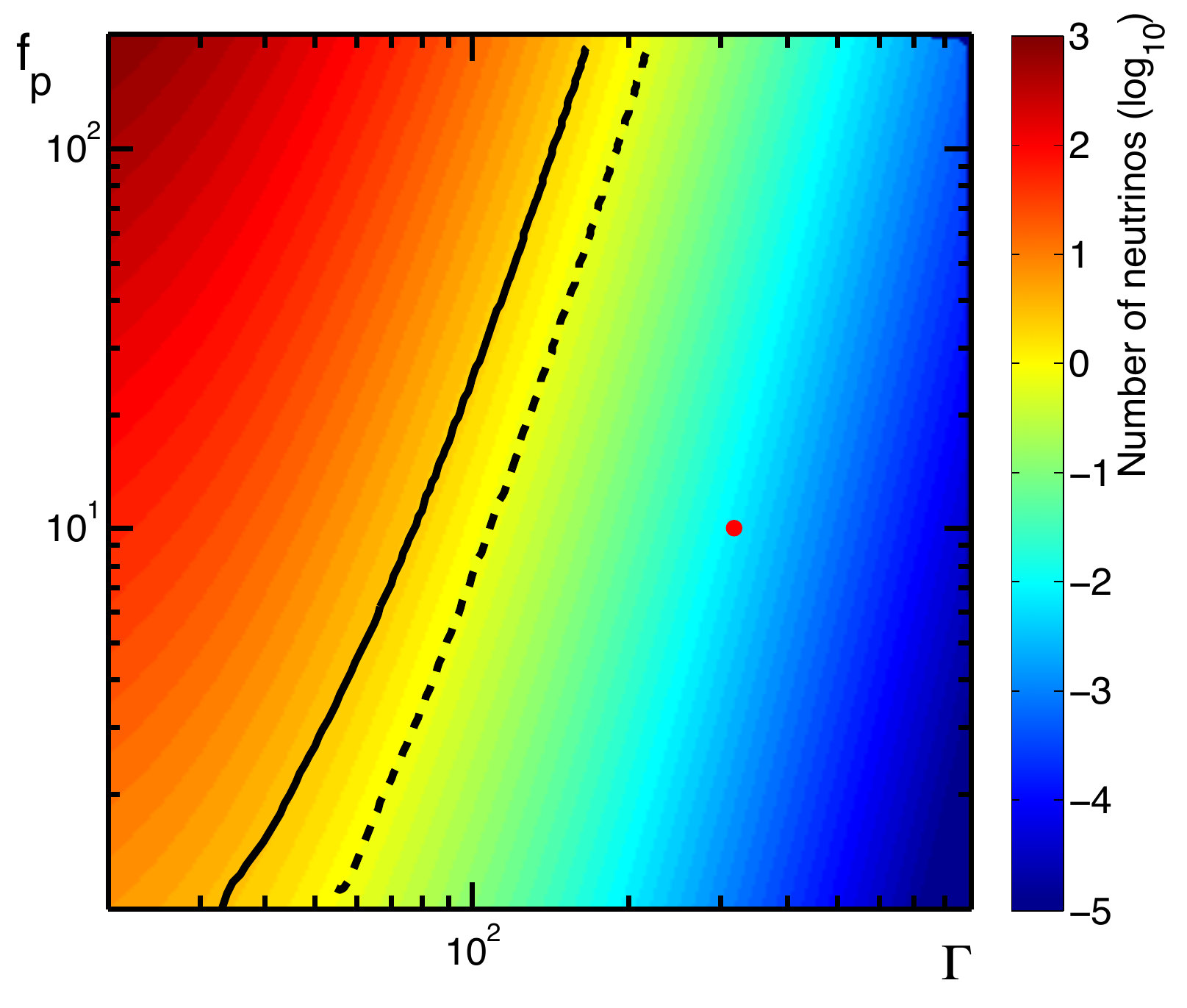}
\end{minipage} \\
\hspace{-1.7cm}
\begin{minipage}{.4\textwidth}
\includegraphics[width=1.25\textwidth]{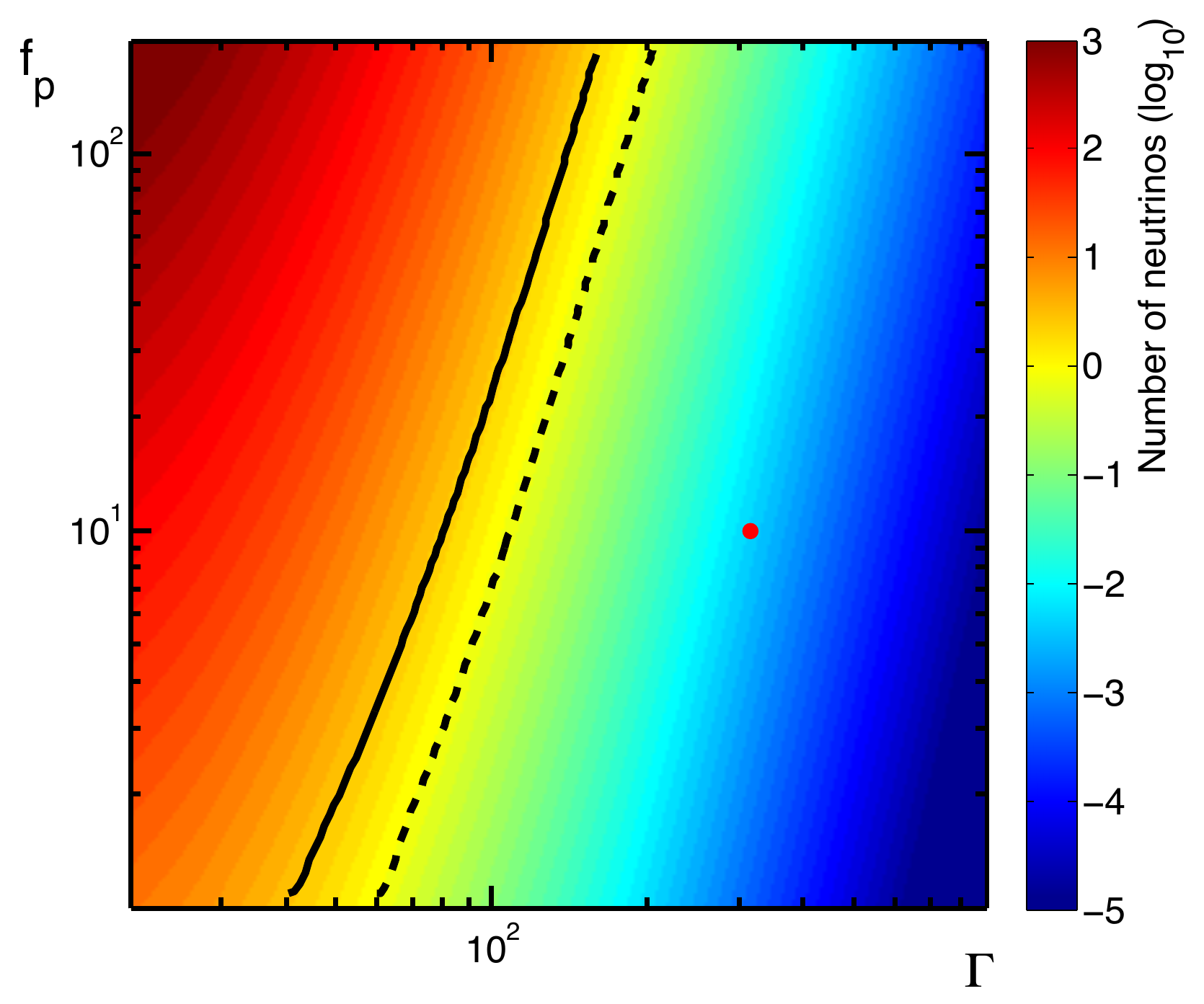}
\end{minipage}
\qquad \hspace{1.3cm}
\begin{minipage}{.4\textwidth}
\includegraphics[width=1.25\textwidth]{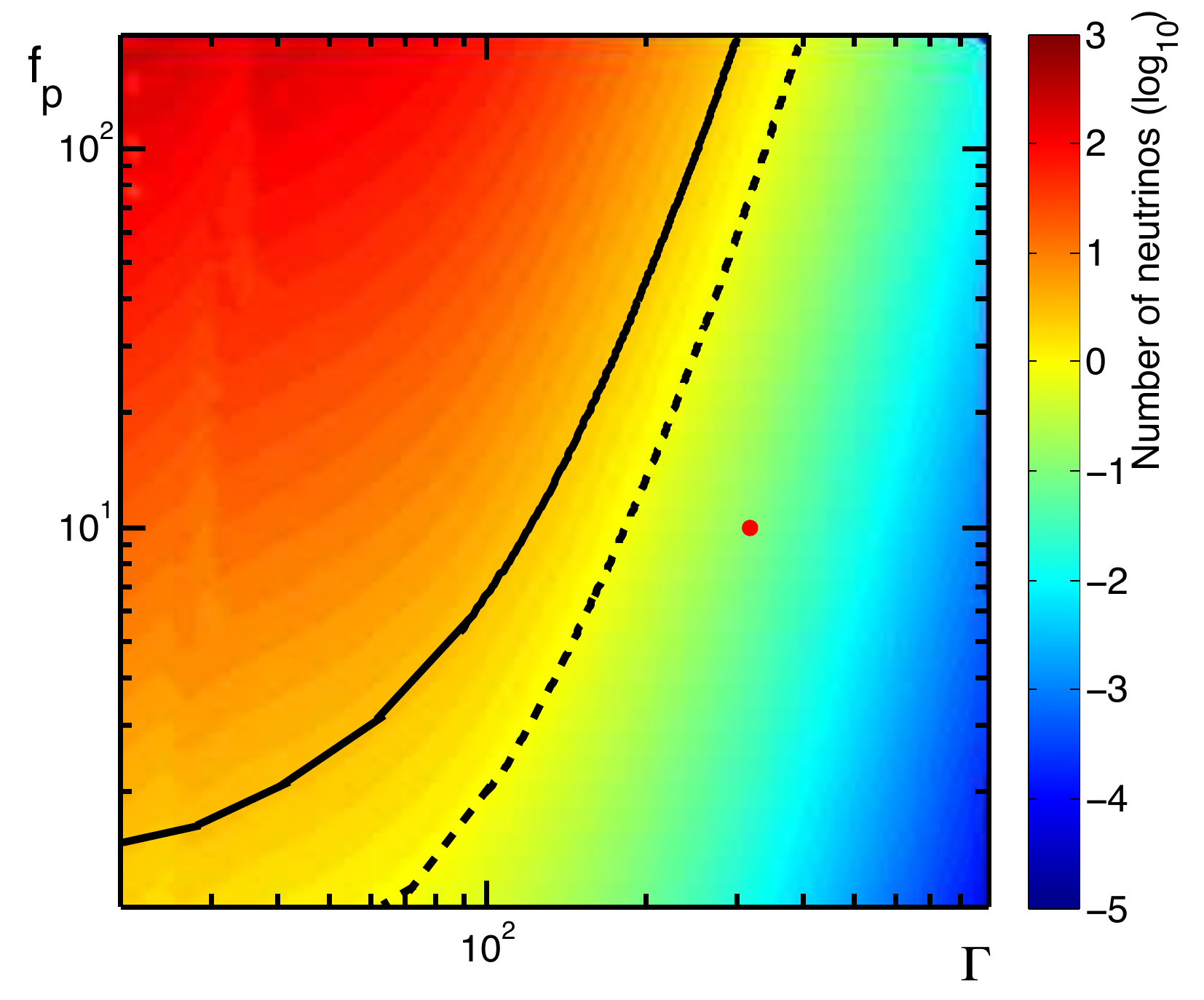}
\end{minipage}
\caption{Number of expected neutrino events detectable with the ANTARES telescope (colored scale) computed as a function of $\Gamma$ and $f_p$, in the context of the IS model. The solid (dashed) black line corresponds to the exclusion limits at 90 (50)$\%$ C.L. The red dot shows the benchmark value $f_p = 10$ and $\Gamma=316$. {\it Top left:} GRB 080916C. {\it Top right:} GRB110918A. {\it Bottom left:} GRB 130427A. {\it Bottom right:} GRB 130505A.} 
\label{fig:constraintsIS}
\end{figure*}

\begin{figure*}
\hspace{-1.7cm}
\begin{minipage}{.4\textwidth}
\includegraphics[width=1.25\textwidth]{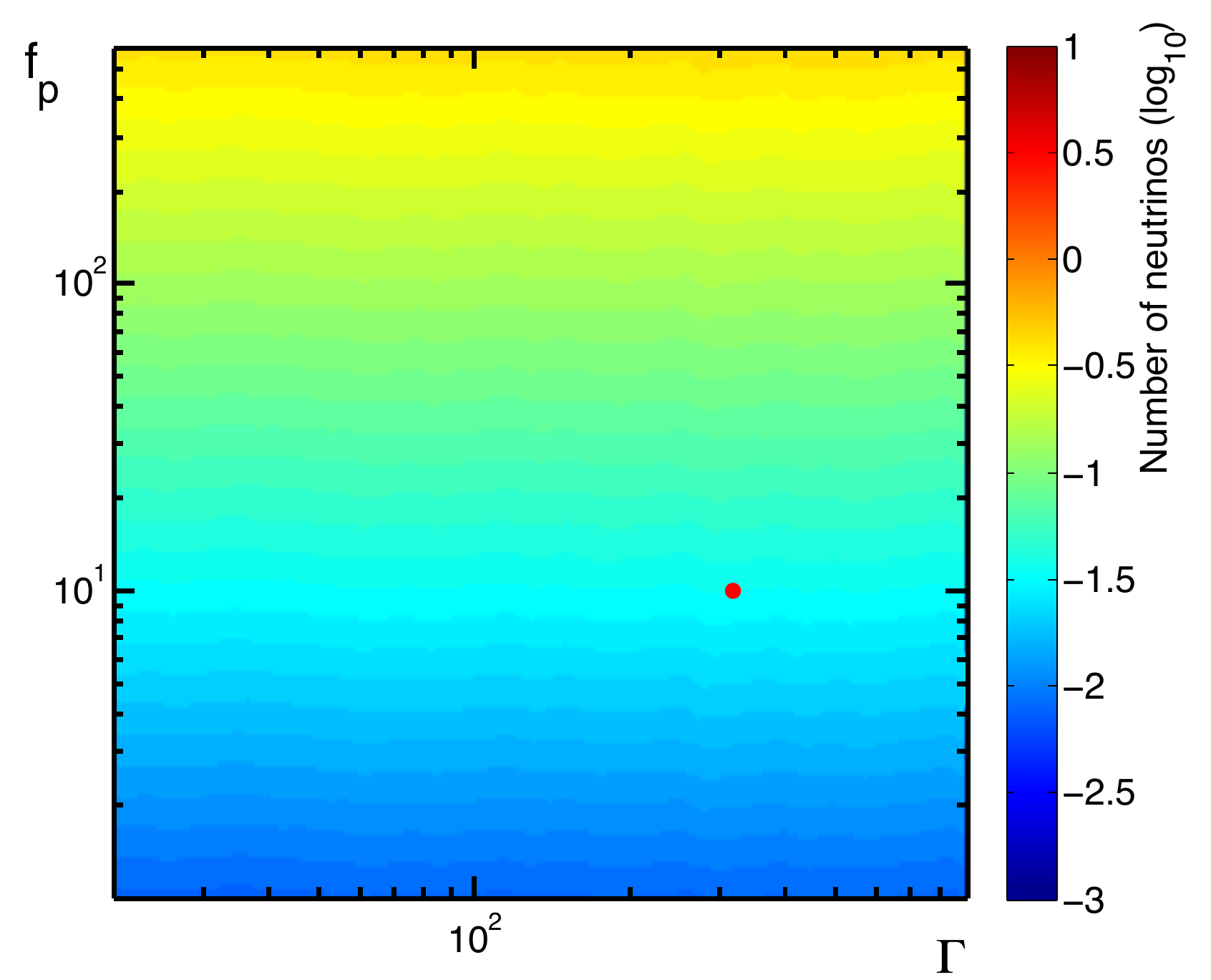}
\end{minipage}
\qquad \hspace{1.3cm}
\begin{minipage}{.4\textwidth}
\includegraphics[width=1.25\textwidth]{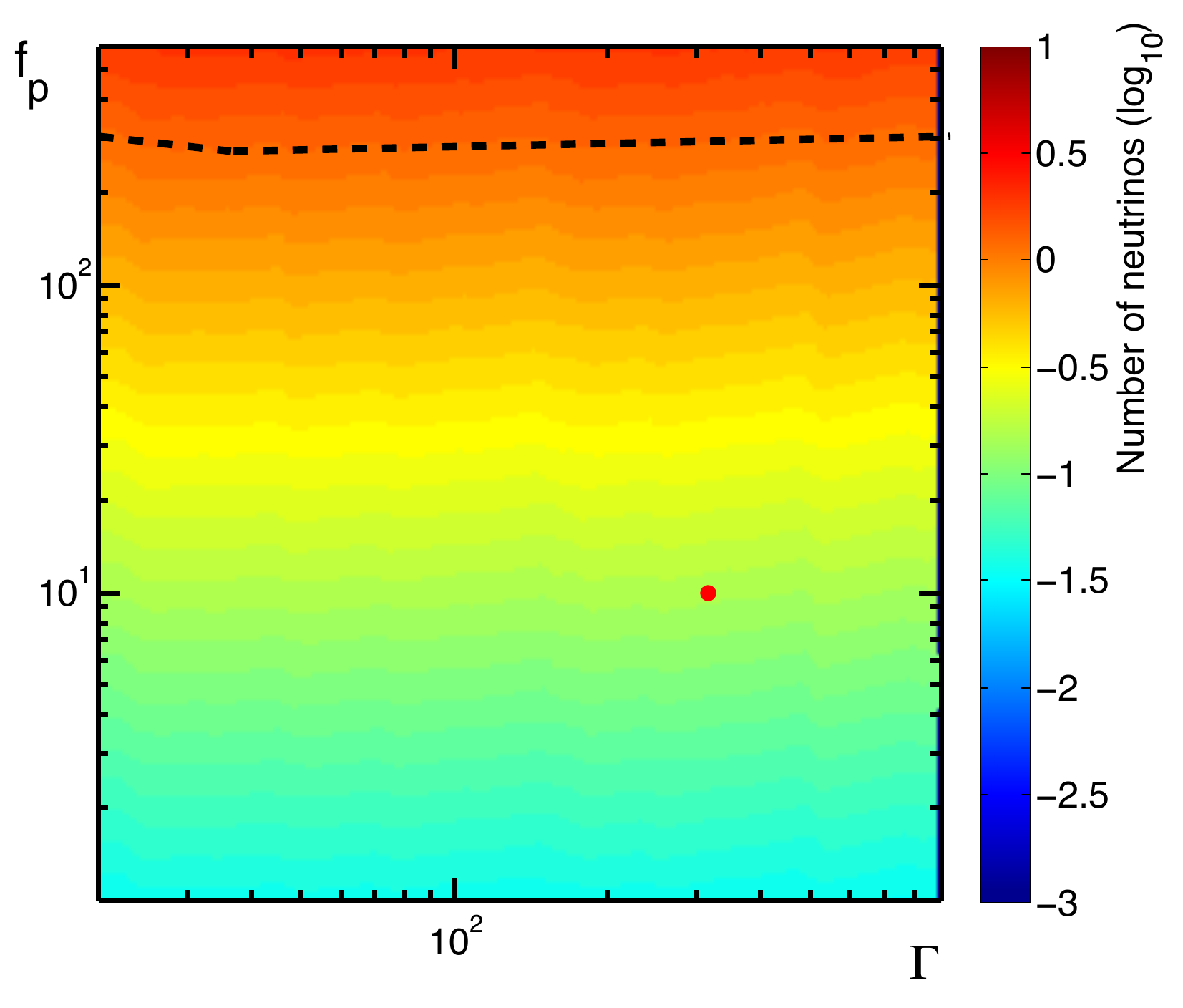}
\end{minipage} \\
\hspace{-1.7cm}
\begin{minipage}{.4\textwidth}
\includegraphics[width=1.25\textwidth]{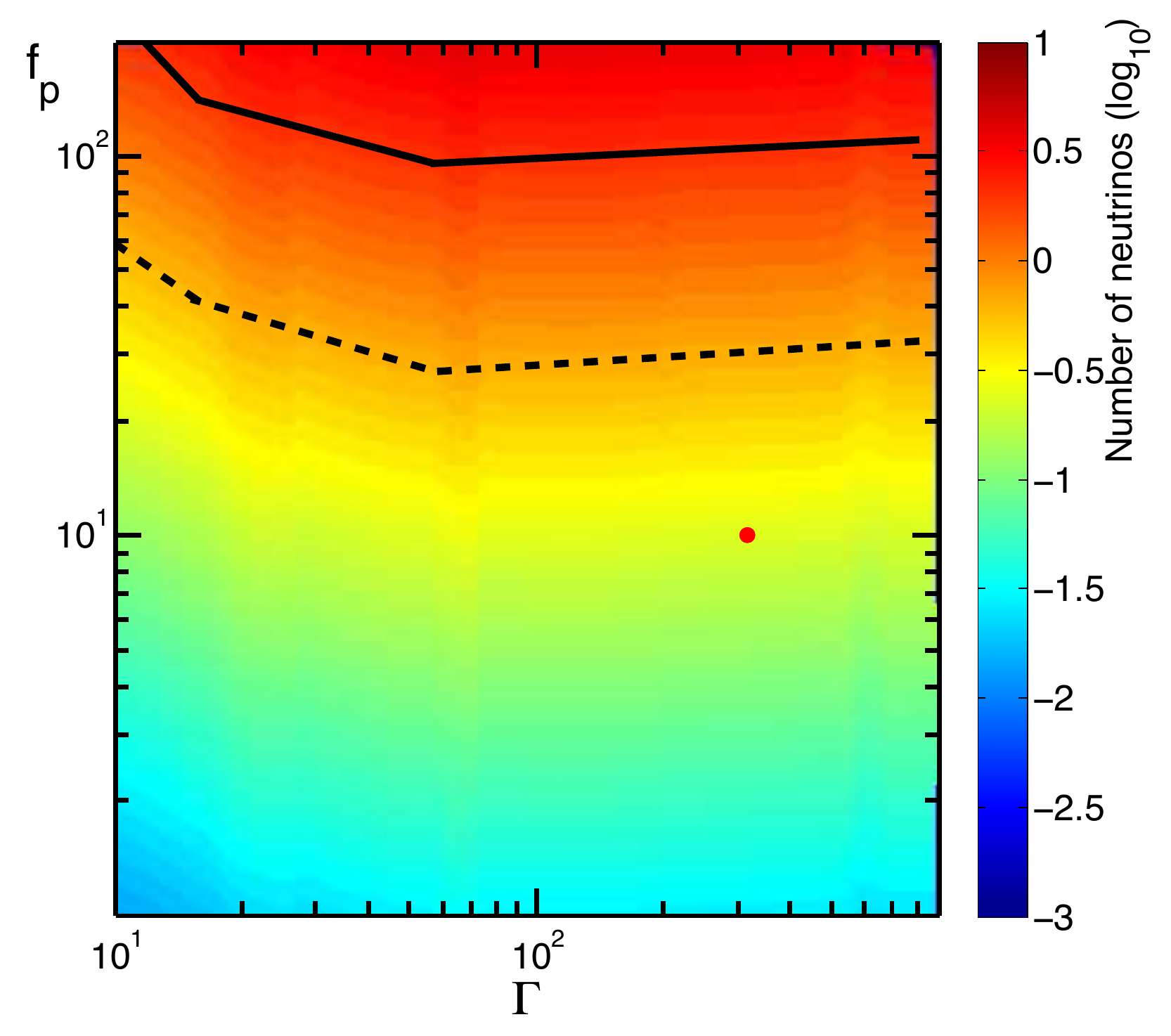}
\end{minipage}
\qquad \hspace{1.3cm}
\begin{minipage}{.4\textwidth}
\includegraphics[width=1.25\textwidth]{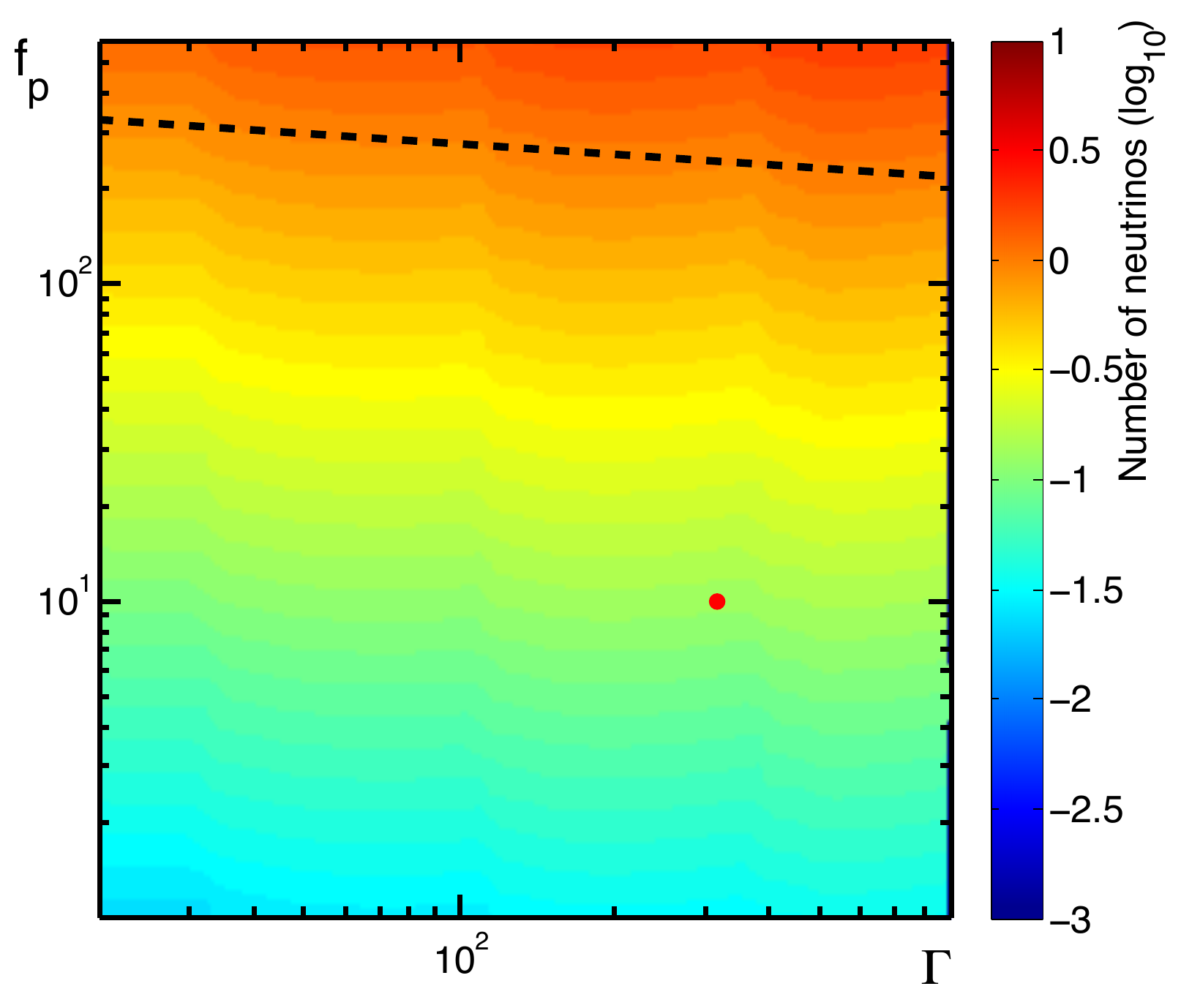}
\end{minipage}
\caption{Number of expected neutrino events detectable with the ANTARES telescope (colored scale) computed as a function of $\Gamma$ and $f_p$, in the context of the PH model. The solid (dashed) black line corresponds to the exclusion limits at 90 (50)$\%$ C.L. The red dot shows the benchmark value $f_p = 10$ and $\Gamma=316$. {\it Top left:} GRB 080916C. {\it Top right:} GRB110918A. {\it Bottom left:} GRB 130427A. {\it Bottom right:} GRB 130505A.}
\label{fig:constraintsPH}
\end{figure*}

The obtained 90$\%$ C.L. limits on the neutrino fluence allow the free parameters that significantly impact the neutrino flux to be constrained both in the framework of the IS and PH model. Since the measured $\gamma$-ray fluence $F_{\gamma}$, the bulk Lorentz factor $\Gamma$ and the baryonic loading factor $f_p$ mainly affect the neutrino yield from GRBs, the use of bright GRBs is justified when assuming that such sources have broadly similar values of $\Gamma$ and $f_p$. However, it is also essential to constrain the much larger sample of faint sources, since they could contribute to the diffuse neutrino flux with their cumulative effect.
In Fig.~\ref{fig:constraintsIS} and \ref{fig:constraintsPH} the 90$\%$ and 50$\%$ C.L exclusion limits are shown in the $\Gamma-f_p$ plane regarding the IS model predictions for all GRBs in the case of the IS and PH models, respectively. It is assumed that $1 \leq f_p \leq 200$ and $10 \leq \Gamma \leq 900$ and that the two parameters are not correlated.

\subsection{Internal Shock Model Case}
For the high-$z$ burst GRB 080916C the derived constraints do not significantly challenge the internal shock model since values of $\Gamma$ above 100 cannot be excluded. At low Lorentz factor regime $\Gamma <100$, values of $f_p$ in the range from 10 to about 30 are excluded but do not go beyond the default value of $f_p$. In the case of this GRB, the constraints are strongly limited because of the large distance to the source. \\
For the two bursts closest to the Earth GRB 130427A ($z=0.34$) and GRB 110918A ($z=0.98$) more stringent limits can be inferred. Low relativistic jets $\Gamma<50$ are completely excluded and a baryonic loading factor is highly constrained to $10 < f_p < 20$ for $50<\Gamma<100$. For $100<\Gamma<200$ values of $f_p$ greater than its benchmark value are excluded, while in the region with $\Gamma>200$ $f_p$ is barely constrained . \\
The most severe constraints are derived for GRB 130505A, starting to significantly challenge the IS scenario up to $\Gamma \sim 200$. This occurs mainly because GRB 130505A is much more energetic than GRB 130427A. In addition, because a short variability time scale was assumed (see Tab.~\ref{tab:param}), its internal shock radius ($R_{\textrm{IS}} \propto t_{\textrm{var}}$) is much smaller (which means that the p$\gamma$ optical depth is enhanced) than that of GRB 110918A. However, contrary to GRB 110918A and GRB 130427A, this burst is much farther away ($z=2.27$) which explains the poorest constraints on $f_p$ at the very low $\Gamma$ regime. Using a different value for the variability time scale, as $ t_{\textrm{var}}=0.1$~s, the constraints are less restrictive and become of the same order of those from GRB 110918A and GRB 130427A.

\subsection{Photospheric Model Case}
The photospheric model is \silvia{not sensitive to the bulk Lorentz factor because of the fact that in correspondence of the photosphere the optical depth of p$\gamma$ interaction is greater than unity and therefore does not depend anymore on $\Gamma$}. Thus the neutrino spectrum is mainly affected by the $\gamma$-ray fluence (and distance effects) and the baryonic loading factor of the sub-photospheric jet. For these reasons less stringent constraints on $f_p$ could be derived for GRB 130505A, GRB 080916C and GRB 110918A. For what concerns GRB 130427A, the closest and the most fluent burst, a high baryonic content (i.e. $f_p>100$) in its jet has been ruled out.

\section{Conclusions}
\label{sec:concl}
A search for muon neutrinos in spatial and temporal coincidence with the prompt emission of four bright GRBs has been performed using ANTARES data. Events satisfying the optimised selection criteria have been considered in two independent analyses, with the purpose to test and constrain the parameters of both the internal shock and the photospheric scenarios of the fireball model. Concerning the internal shock model, the analysis has been optimised in order to give the highest model discovery potential for each burst, relying on the numerical model NeuCosmA. For the photospheric model the search strategy has been adapted using a dedicated data sample, able to enhance the sensitivity of the detector in the neutrino energy range between 100~GeV and 1~TeV, and optimised in the same way. No signal events have been detected in any of the searches, so that 90$\%$ C.L. upper limits on $E_{\nu}^2 \phi_{\nu}$ are derived. For the internal shock model, they are placed between $10^{-1}$~GeV/cm$^2$ and $10$~GeV/cm$^2$ in the neutrino energy range going from $3 \times 10^{4}$~GeV to $2 \times 10^{7}$~GeV. For the photospheric model they stand in the same interval, but in the lower neutrino energy range from $1 \times 10^{2}$~GeV to $3 \times 10^{4}$~GeV. This search extends the ANTARES neutrino detection capability from GRBs into the low-energy regime; compared to what was shown in previous ANTARES searches for muon neutrinos in coincidence with 296 GRBs during four years of data (\citealt{Julia}), it also confirms the sensitivity in the high-energy regime, i.e. above 100~TeV. Existing limits cannot rule out the theoretical models investigated here. It is worth recalling, however, that the expected neutrino fluence is normalised to the detected $\gamma$-ray emission: this allows to constrain the parameters affecting the GRB emission mechanism. In particular, limits on the bulk Lorentz factor and on the baryonic content of the GRB jet according to the IS/PH scenarios have been derived for each source. Assuming the internal shocks, for the closest burst the results suggest a low neutrino production efficiency because of the high $\Gamma$ region still allowed. Such a picture is supported by the Lorentz factor estimation performed for the selected energetic bursts: $\Gamma = 870$ for GRB 080916C (\citealt{latgbm}), $\Gamma = 340-450$ for GRB 130427A (\citealt{Hascoet2015} and \citealt{Vurm2015}) and $\Gamma = 340$ for GRB 110918A (\citealt{frederiks}). This fact may work against the detection of high-energy neutrinos: the high neutrino production expected in the jet of the most fluent GRBs seems to be compensated by a high Lorentz factor and possibly by a low baryonic loading.
Models that assume that a low fraction of the GRB kinetic energy is transferred to protons (low $f_p$) if  $\Gamma$ is high are the most difficult to constrain using neutrino telescopes, as evident from both Fig.~\ref{fig:constraintsIS} and \ref{fig:constraintsPH}. The constraints do not exclude the hypothesis that, for a given jet energy, high values of $\Gamma$ imply small values of $f_p$, as suggested by \citealt{Sari1995}. This effect (low $f_p$ if $\Gamma$ is high) goes against the intuitive idea that the most energetic bursts (and generally the most fluent ones) are the best targets for individual neutrino detection. 
In the case of the photospheric scenario, on the other hand, less stringent constraints could be placed and most of the parameter space is still available. \\
The same constraints can in principle provide information on the allowed energy range and on the composition of primary particles. The connection between constraints in neutrinos and CR measurements indicates that a multi-messenger approach is a suitable strategy in the framework of testing the paradigm of GRBs as UHECR sources. Current neutrino telescopes have a small probability to detect neutrinos from GRBs, as shown in Tab.~\ref{tab:optimum}: further investigations of this scenario will be possible with the incoming generation of neutrino detectors, such as KM3NeT-ARCA (\citealt{km3}) and IceCube-GEN2 (\citealt{gen2}).

\section*{Acknowledgements}
The authors acknowledge the financial support of the funding agencies: Centre National de la Recherche Scientifique (CNRS), Commissariat \`a l'\'ener\-gie atomique et aux \'energies alternatives (CEA), Commission Europ\'eenne (FEDER fund and Marie Curie Program), Institut Universitaire de France (IUF), IdEx program and UnivEarthS Labex program at Sorbonne Paris Cit\'e (ANR-10-LABX-0023 and ANR-11-IDEX-0005-02), Labex OCEVU (ANR-11-LABX-0060) and the A*MIDEX project (ANR-11-IDEX-0001-02), R\'egion \^Ile-de-France (DIM-ACAV), R\'egion Alsace (contrat CPER), R\'egion Provence-Alpes-C\^ote d'Azur, D\'e\-par\-tement du Var and Ville de La Seyne-sur-Mer, France; Bundesministerium f\"ur Bildung und Forschung (BMBF), Germany; Istituto Nazionale di Fisica Nucleare (INFN), Italy; Stichting voor Fundamenteel Onderzoek der Materie (FOM), Nederlandse
organisatie voor Wetenschappelijk Onderzoek (NWO), the Netherlands; Council of the President of the Russian Federation for young scientists and leading scientific schools supporting grants, Russia; National Authority for Scientific Research (ANCS), Romania;  Mi\-nis\-te\-rio de Econom\'{\i}a y Competitividad (MINECO): Plan Estatal de Investigaci\'{o}n (refs. FPA2015-65150-C3-1-P, -2-P and -3-P, (MINECO/FEDER)), Severo Ochoa Centre of Excellence and MultiDark Consolider (MINECO), and Prometeo and Grisol\'{i}a programs (Generalitat Valenciana), Spain; Agence de  l'Oriental and CNRST, Morocco. We also acknowledge the technical support of Ifremer, AIM and Foselev Marine for the sea operation and the CC-IN2P3 for the computing facilities.








\bsp	
\label{lastpage}
\end{document}